\documentclass[prl,aps,twocolumn,longbibliography]{revtex4-1}
\usepackage{color}
\usepackage{graphicx}
\usepackage{bm}
\usepackage{amsmath}
\usepackage{amssymb}
\usepackage{amsthm}
\usepackage{amsfonts}
\usepackage{enumerate}
\usepackage{latexsym}
\usepackage{hyperref}

\begin{document}

\title{ Topological magnetotorsional effect in Weyl semimetals}
\author{Long Liang$^{1,2}$ and  Teemu Ojanen$^{2,3}$}
\affiliation{1, Department of Applied Physics, Aalto University School of Science, FI-00076 Aalto, Finland \\2, Computational Physics Laboratory, Physics Unit, Faculty of Engineering and
Natural Sciences, Tampere University, P.O. Box 692, FI-33014 Tampere, Finland}
\affiliation{3, Helsinki Institute of Physics P.O. Box 64, FI-00014, Finland}

\begin{abstract}
In this work we introduce a thermal magnetotorsional  effect (TME) as a novel topological response in magnetic Weyl semimetals. We predict that  magnetization gradients perpendicular to the Weyl node separation give rise to temperature gradients depending only on the local positions of the Weyl nodes. The TME is a consequence of magnetization-induced effective torsional spacetime geometry and the finite temperature Nieh-Yan anomaly. Similarly to anomalous Hall effect and chiral anomaly, the TME has a universal material-independent form. We predict that the TME can be observed in magnetic Weyl semimetal EuCd$_2$As$_2$.
\end{abstract}
\maketitle

\emph{Introduction-- } The cross-fruition between condensed matter and high-energy physics has become a central theme in contemporary research. 
Topological Weyl and Dirac semimetals~\cite{RevModPhys.90.015001} 
provide a particularly rich example of this interplay. These systems offer a versatile playground to study phenomenology of relativistic fermions in the presence of gauge fields and curved spacetimes as well as inspire novel electronic applications.  Due to the flexibility of condensed matter systems, there are no fundamental limitations to realize predictions that have remained hypothetical in the high-energy context. A fascinating example of such predictions is the Nieh-Yan anomaly~\cite{Chandia_1997,Onkar_2014,nissinen2019emergent,NissinenVolovik_2019,NissinenVolovik_2019_2,Huang_2019,Huang_2019_2} of chiral fermions in torsional spacetimes which has recently been proposed as a source of universal effects in condensed matter systems. 

\begin{figure}
	\includegraphics[width=0.45\textwidth]{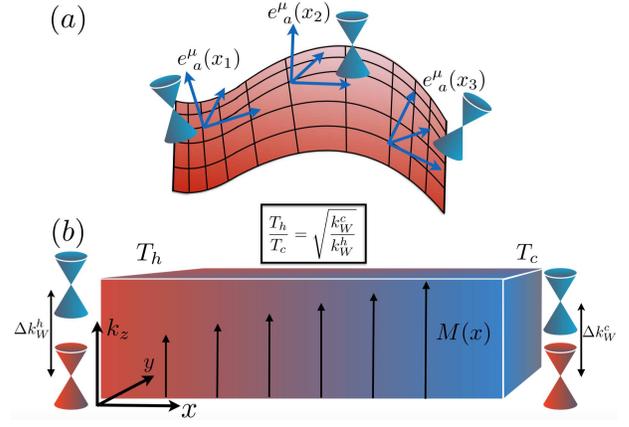}
	\caption{(a) Weyl fermions couple to spacetime geometry through the frame fields $e^\mu_{~a}$. Inhomogeneous magnetization gives rise to locally varying frame fields in Weyl semimetals. (b) Thermal magnetotorsional effect gives rise to temperature gradient as a response to magnetic gradient perpendicular to the Weyl node separation $\Delta k_W$. Placing one end of a sample in contact with a heat bath, the relative temperature drop between the ends is given by the relative shift of the node separation. }\label{Fig:Cartoon}
\end{figure}

Significant interest in topological semimetals is focused on their remarkable material-independent properties resulting from momentum-space Berry monopoles and field theory anomalies. In the present work, we report a discovery of a novel addition to the previously known topological responses in Weyl semimetals. We introduce the thermal magnetotorsional effect (TME) which gives rise to energy currents and temperature gradients as a response to magnetic gradients. This effect results from two important factors. First of all, as depicted in Fig.~\ref{Fig:Cartoon} (a), the Weyl fermions couple to spacetime geometry through frame fields. As established in Refs.~\cite{WeylMetamaterials_PRX2017,LiangOjanen_2019}, smoothly varying magnetization will give rise to a low-energy description in terms of locally varying frame fields. As discussed in the present work, the effective geometry encoded in frame fields admit non-vanishing torsion. Another key ingredient is the Nieh-Yan anomaly of chiral fermions in torsional spacetimes. Until recently, physical consequences of the Nieh-Yan anomaly~\cite{Chandia_1997,Onkar_2014} have remained controversial due to the explicit appearance of the non-universal high-energy cutoff in physical predictions. However, several recent studies~\cite{NissinenVolovik_2019,NissinenVolovik_2019_2,Huang_2019,nissinen2019emergent} have confirmed its universal character by identifying temperature as the appropriate substitute for the high-energy cutoff.

We first present a derivation of an effective low-energy theory of magnetic Weyl materials with torsional geometry. We  proceed by showing how the TME results from combining the effective geometry with the finite temperature Nieh-Yan anomaly. Then we discuss the striking physical consequences of the TME by studying a system depicted in Fig.~\ref{Fig:Cartoon} (b). By connecting a magnetic Weyl material to thermal bath at one end, the relative temperature drop in the system is given by the relative shift of the node separation $T_h/T_c=\sqrt{k_W^c/k_W^h}$. This remarkable relation has the same universal form in all Weyl materials and is independent on the specific spatial profile of the magnetic texture. We discuss how this relation can be experimentally probed in magnetic Weyl material EuCd$_2$As$_2$~\cite{Wang_2019,Soh_2019,Ma_2019}. 

\emph{Torsional spacetimes in magnetic Weyl semimetals-- } 
We first outline the general mechanism for the emergent torsional geometry in time-reversal (TR) breaking Weyl semimetals. We consider a four-band parent Hamiltonian that breaks the TR symmetry 
\begin{eqnarray}\label{model}
H=k_i\gamma_i + m\gamma_4+\mathbf{M}(\mathbf{r})\cdot \mathbf{b},
\end{eqnarray}
where $\gamma_i$ with $i=1, 2, 3, 4$ denotes the four Hermitian $\gamma$ matrices satisfying anticommutation relations $\{ \gamma_i, \gamma_j\}=2\delta_{ij}$ with $\delta_{ij}$ being the Kronecker delta function and $\mathbf{b}=(\gamma_{23}, \gamma_{31}, \gamma_{12})$ with $\gamma_{ij}=-i[\gamma_i, \gamma_j]/2$. 
Since the momentum is odd under the TR, $\gamma_{1,2,3}$ and $\mathbf{b}$ are also odd. The $\mathbf{M}(\mathbf{r})\cdot \mathbf{b}$ term breaks the TR symmetry, and  $\mathbf{M}(\mathbf{r})=M(\mathbf{r})(\sin{\theta(\mathbf{r})}\cos{\phi(\mathbf{r})}, \sin{\theta(\mathbf{r})}\sin{\phi(\mathbf{r})}, \cos{\theta(\mathbf{r})})$  corresponds to 3-dimensional (3d) magnetization or any field that transforms as magnetization under TR. It can be used to describe  real materials such as 3d topological insulators (TI) with magnetic texture~\cite{Cho_2011}, topological insulator heterostructures~\cite{PhysRevLett.107.127205,PhysRevB.84.235126} or ferromagnetic Weyl semimetals discovered recently~\cite{Belopolski1278,Liu1282,Morali1286,Wang_2019,Soh_2019,Su_2019,Ma_2019}. 

Following the general method introduced in Refs.~\cite{WeylMetamaterials_PRX2017,LiangOjanen_2019}, we block diagonalize the parent Hamiltonian [for details, see the supplemental information (SI)]. This leads to an effective Weyl Hamiltonian, 
\begin{eqnarray}\label{Eq:Weyl}
H_W&=& d_a(\mathbf{k},\mathbf{r})\sigma^a,
\end{eqnarray}
where $a=0,1,2,3$, $\sigma^0$ is the $2\times 2$ unit matrix, $\sigma^{1,2,3}$ are the Pauli matrices,  $d_1=\cos{\theta}\cos{\phi} k_x+\cos{\theta}\sin{\phi} k_y-\sin{\theta}k_z$, $d_2=-\sin{\phi}k_x+\cos{\phi}k_y$, $d_3=M-m-(\kappa^2_3+f^2)/2m$, $d_0=-f \kappa_3 /m$ with $\kappa_3=\cos{\phi}\sin{\theta}k_x+\sin{\phi}\sin{\theta}k_y+\cos{\theta}k_z$, and $f=(\partial_z\phi + \cos{\phi}\partial_y\theta-\sin{\phi}\partial_x\theta)/2$. The Weyl points are located at $\pm \mathbf{K}_W=\pm K_W(\sin{\theta}\cos{\phi}, \sin{\theta}\sin{\phi},\cos{\theta})$ with $K_W=\sqrt{2m(M-m)-f^2}$. We assume  $2m(M-m)>f^2$ such that there are always two well separated Weyl points.

Expanding Eq.~\eqref{Eq:Weyl} around the Weyl point $\mathbf{K}_W$, we obtain the Hamiltonian for the left-handed Weyl fermion
\begin{eqnarray}\label{Eq:Weyl_lowenergy}
H_L \approx \frac{1}{2}\{v^i_{~a} ,k_i-K_{W,i}\}\sigma^a -K_{W,0}\sigma^0, 
\end{eqnarray}
where $K_{W,0}=fK_W/m$ is the energy shift of the Weyl point and $v^i_{~a}(\mathbf{r})=\frac{\partial d_a}{\partial k_i}\big|_{\mathbf{K}_W}$ is position dependent Fermi velocity. It is convenient to introduce $v^{0}_{~0}=-1$ and $v^{0}_{~1,2,3}=0$ such that $v$ can be written as a $4\times 4$ matrix.
The explicit expressions for $v^\mu_{~a}$ are given in SI. The effective Weyl Hamiltonian is similar to the one induced by elastic deformations~\cite{Cortijo_2016}, where the position dependent Fermi velocities are related to the strain instead of magnetization.  However, an important difference is that the space-time mixed components, $v^i_{~0}$, are absent in the strained Weyl systems, whereas in inhomogeneous magnetic Weyl semimetals the mixed term can exist and give rise to exotic physics~\cite{LiangOjanen_2019}.

The action corresponding to the left-handed Weyl fermion takes the form (see SI)
\begin{eqnarray}\label{Eq:S_Weyl_left}
	S_L&=&\int\mathrm{d}^4x~e\psi^\dag_L e^{\mu}_{~a}\left(i\partial_\mu+A_{5,\mu}+i \frac{T_{\mu}}{2}\right) \sigma^a\psi_L,
\end{eqnarray}
where $\psi_L$ is a two-component spinor, $e\equiv \mathrm{det}(e_\mu^{~a})= (\mathrm{det}\,  v^\mu_{~a})^{1/3}$ gives the invariant volume element, and  $ e^{\mu}_{~a}=e^{-1}v^\mu_{~a}$ is the frame field with the coordinate index denoted by $\mu=t,x,y,z$ and the Lorentz index by $a=0,1,2,3$. The emergent metric is given  $g_{\mu\nu}=e_\mu^{~a}e_\nu^{~b}\eta_{ab}$ with $\eta_{ab}=\mathrm{diag}(-1,1,1,1)$ and $e_\mu^{~a}$ being the coframe field satisfying $e_\mu^{~a} e^\nu_{~a}=\delta^\nu_\mu$ and $e_\mu^{~a} e^\mu_{~b}=\delta^a_b$. 
The position of the Weyl point is given by the ``axial gauge field'' $A_{5,\mu}=K_{W,\mu}$.
Comparing to the standard action of a spinor field in gravitational fields~\cite{2011arXiv1106.2037Y}, the spin connection is absent in Eq.~\eqref{Eq:S_Weyl_left}. As a result, the curvature also vanishes and the torsion two form $T^a$ is simply given by the exterior derivative of the coframe one form $e^a$, $T^a=d e^a$. 
The torsion is in general nonzero and therefore the inhomogenous magnetization induces a low energy Weyl fermion living in a Weitzenb\"ock spacetime, which is used in teleparallel gravity  theory~\cite{PhysRevD.19.3524}. The coframe field $e^a$ can also be viewed as a gauge potential, and then the torsion is the field strength corresponding to the potential.
Note that it is the trace of torsion tensor, $T_{\mu}=T^{~~\nu}_{\mu\nu} =e^{-1}e_{\mu}^{~a} \partial_{\nu}(e e^\nu_{~a})$,  that appears in the action. We consider a static magnetization so the frame fields are time independent, but in general it is possible to realize time-dependent frame fields. Thus, in the following derivation of the TME we will allow for time-dependent frame fields and formulate the general theory in a covariant form.

\emph{Finite temperature mixed chiral-torsional anomaly-- } The low energy theory Eq.~\eqref{Eq:S_Weyl_left} has far-reaching  physical consequences. Here we focus on the torsion-induced effects which have attracted significant attention recently~\cite{Hughes_2011,Hidaka_2012,Hughes_2013,Onkar_2014,NissinenVolovik_2019,NissinenVolovik_2019_2,Huang_2019,Huang_2019_2,You_2014,PhysRevLett.116.166601,Zubkov_2016,Khaidukov_2018,Imaki_2019,PhysRevLett.122.056601,PhysRevB.99.024513,PhysRevB.99.155152}. It has been shown that the torsion gives a contribution to the chiral anomaly~\cite{Chandia_1997} through the Nieh-Yan term~\cite{NiehYan_1982}, which in the absence of spin connection reads 
\begin{eqnarray}
N=\frac{\varepsilon^{\mu\nu\rho\sigma}}{4}T^{~~a}_{\mu\nu}T_{\rho\sigma a}=2\mathbf{E}^a\cdot\mathbf{B}_a,
\end{eqnarray}
where $\varepsilon^{\mu\nu\rho\sigma}$ is the totally antisymmetric symbol  with the convention $\varepsilon^{txyz}=1$, and $\mathbf{E}^a$  and $\mathbf{B}^a$ are the torsional electric and magnetic fields, respectively. 
The Nieh-Yan term looks similar to the celebrated Adler-Bell-Jackiw anomaly~\cite{BellJackiw_1969,Adler_1969}.
However, a fundamental difference is that the Nieh-Yan term has the dimension of inverse length squared $[L^{-2}]$ while the electromagnetic  $\mathbf{E}\cdot\mathbf{B}$ term has  the dimension of $[L^{-4}]$. Consequently, the coefficient of the Nieh-Yan contribution to the chiral anomaly has the dimension of  $[L^{-2}]$. It was pointed out that the coefficient depends on the high-energy cutoff in Weyl semimetals~\cite{Onkar_2014} and therefore makes the contribution superficially nonuniversal.  Recently, it was suggested~\cite{NissinenVolovik_2019,NissinenVolovik_2019_2} that the temperature can also play the role of cutoff since it has the correct dimension, and the temperature dependent contribution of Nieh-Yan anomaly takes a universal form
\begin{eqnarray}
\partial_{\mu}(e J^{\mu}_5)=\frac{T^2}{12}N.
\end{eqnarray}
This result has been verified by a direct calculation of the chiral charge density induced by torsional Landau levels, establishing the Nieh-Yan anomaly as a universal Fermi-surface effect at finite-temperatures. Furthermore, it has been argued that in addition to temperature, coefficient contains dimensionless central charge of (1+1)-d Dirac fermions~\cite{Huang_2019}.  Since the central charge is also related to the (1+1)-d gravitational anomaly~\cite{Bradlyn_2015}, there could be a connection between the thermal Nieh-Yan anomaly in (3+1)-d and the gravitational anomaly in (1+1)-d which is yet to be revealed, moreover, temperature dependent torsional anomalies may also appear in other dimensions~\cite{Huang_2019_2}. However, the most important insight from the recent activity is the emergence of a simple physical picture of the Nieh-Yan contribution to the chiral anomaly in terms of torsional Landau levels and its universal nature at finite temperature.

From the thermal Nieh-Yan anomaly, we find a contribution to the effective action
\begin{eqnarray}\label{Eq:S_anomaly}
S_{\mathrm{anomaly}}=\frac{T^2}{12}\int \mathrm{d}^4 x~ \varepsilon^{\mu\nu\rho\sigma} A_{5\mu}e_{\nu}^{~a}\partial_\rho e_{\sigma a}.
\end{eqnarray}
This effective action leads to a  chiral current $ej^{\mu}_5=T^2\varepsilon^{\mu\nu\rho\sigma} e_{\nu}^{~a}\partial_\rho e_{\sigma a}/12$, which is consistent with the current obtained from linear response calculations~\cite{Khaidukov_2018,Imaki_2019} and therefore provides additional confirmation on the validity of the anomaly action. The original action, Eq.~\eqref{Eq:Weyl_lowenergy}, exhibits the Nieh-Yan-Weyl rescaling symmetry~\cite{Hughes_2013} $e^a\to \exp{(\Lambda)} e^a$ which does not carry over to the quantum theory because of the anomaly. The overall normalization of the frame fields in the effective action is fixed by requiring that the chiral densities resulting from the direct torsional Landau level calculation is equal to the one obtained from the effective action. Thus we find that the frame fields should be normalized so that the volume element is unity $e=1$, see the SI for details. The zero temperature counterpart of Eq.~\eqref{Eq:S_anomaly} was discussed in Refs.~\cite{Hughes_2011,Hughes_2013,Onkar_2014,You_2014} in the context of lattice dislocations. However, due to the cutoff-dependent and impractical nature of the physical consequences, the experimental signatures of the Nieh-Yan anomaly have remained unclear. In contrast, as shown below, the finite temperature anomaly gives rise to profound universal effects that can be observed experimentally.
 
\emph{Topological magnetotorsional effect--} Taking variation of the effective action Eq.~\eqref{Eq:S_anomaly} with respect to the coframe fields, we obtain the anomalous energy-momentum current
\begin{eqnarray}\label{Eq:Energy_Mometum}
e T^{\mu}_{~a}=-\frac{T^2}{6} \varepsilon^{\mu\nu\rho\sigma} A_{5\nu}\partial_\rho e_{\sigma a}+\frac{T^2}{12} \varepsilon^{\mu\nu\rho\sigma} \partial_\nu A_{5\rho} e_{\sigma a}.
\end{eqnarray}
The second term on the right-hand side of the above equation is omitted in~\cite{Huang_2019}, but in our case it is of the same order as the first term since the frame fields and axial gauge potential have the same physical origin. It can also be understood as the 0th chiral pseudo-Landau level contribution to the energy current~\cite{Landsteiner_2016}.
For $a=0$, we get the energy current. The physical energy-momentum current corresponds to $eT^{\mu}_{~\nu}$, which is obtained by converting the lower Lorentz index to the spacetime index with the help of coframe fields~\cite{Bradlyn_2015_0}, but this does not affect our main result, so we use $e T^{\mu}_{~a}$ for simplicity. 

Assuming $A_{5,t}=0$, then the energy current reads ($J^i_{\varepsilon} \equiv e T^{i}_{~0}$)
\begin{eqnarray} \label{Je}
\mathbf{J}_{\varepsilon} = -\frac{T^2}{6}\mathbf{A}_5\times \mathbf{E}_0-\frac{T^2}{12}\mathbf{B}_5 e_{t0},
\end{eqnarray}
where $\mathbf{B}_5=\nabla \times \mathbf{A}_5$ is the chiral magnetic field and $E_{i,0}=\partial_i e_{t0}-\partial_t e_{i0}$ is the torsional electric field. Remarkably, since the chiral vector potential   $\mathbf{A}_5=(k_W/M)\mathbf{M}(\mathbf{r})$ and the frame fields $e_{\mu0}$ are directly given in terms of the local magnetization $\mathbf{M}(\mathbf{r})$, Eq.~\eqref{Je} in fact expresses the energy current as a response to magnetization. This gives rise to novel magnetothermal response in Weyl semimetals.  We call relation~\eqref{Je}, \emph{combined with} the expressions for magnetization-induced $\mathbf{A}_5$ and $e_{\mu0}$, as the TME. The energy current is perpendicular to the separation between the Weyl points.

On the other hand, the temperature gradient is also a driving force of the energy current. Temperature gradients can be formally identified with an extra torsional electric field $T^{-1}\nabla T=-\mathbf{E}_0$~\cite{Shitade_2014,Tatara_2015,Nakai_2017}. As a direct consequence of this prescription, the anomalous thermal Hall current is given by $\mathbf{J}_\epsilon = T\mathbf{A}_5\times \nabla T/6$~\cite{Huang_2019,Landsteiner_2016}. In the equilibrium, the energy current vanishes~\cite{PhysRevLett.123.060601} and Eq.~\eqref{Je} implies that
\begin{eqnarray}\label{TME}
\mathbf{A}_5\times\frac{\nabla T}{T}=\frac{1}{2}\nabla\times \mathbf{A}_5.
\end{eqnarray}
This relation, coupling magnetization $\mathbf{A}_5=(k_W/M)\mathbf{M}(\mathbf{r})$ to thermal gradients, follows from Eq.~\eqref{Je} together with the expressions for magnetization-induced gauge and frame fields and is the main result of this paper. It presents a new type of response with topological origin and concrete observable effects.

To demonstrate the remarkable consequences of the TME, we consider a magnetic Weyl semimetal model, Eq.~\eqref{model} with unidirectional magnetization $\mathbf{M}=(0,0,M_3(x))$. In this case the Weyl node separation is parallel to $\mathbf{M}$ and Eq.~\eqref{TME} gives rise to the temperature profile
\begin{eqnarray}\label{Eq:Temperature}
\frac{T(x)}{T(0)}=\sqrt{\frac{K_W(0)}{K_W(x)}},
\end{eqnarray}
where $K_W(x)=\sqrt{2m(M_3(x)-m)}$. The relation Eq.~\eqref{Eq:Temperature} has a universal form and the material specific details only enter through the dependence of Weyl node positions on magnetization. In this simple model, the magnetization-induced torsional electric field vanishes, but it is generally nonzero for time-dependent magnetization or interacting Weyl semimetals~\cite{Nissinen_2017}. As shown in the SI, by replacing the node separation by its component perpendicular to the temperature gradient $k_W\to k_{W\perp}$, the relation \eqref{Eq:Temperature} also applies to rotating N\'eel-type magnetic textures such as the one depicted in Fig.~2 (a).

\begin{figure}
	\includegraphics[width=0.4\textwidth]{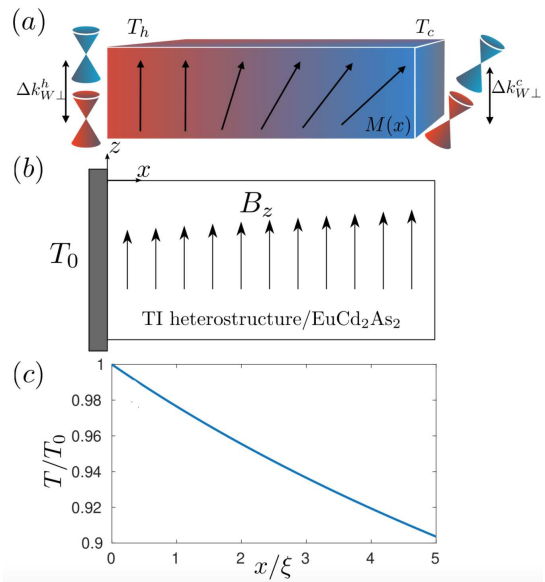}
	\caption{(a): N\'eel-type textures in magnetic Weyl semimetals will induce temperature gradients. (b): A spatially varying magnetic field is applied in the $z$ direction, and the temperature at $x=0$ is fixed by a thermal bath. (c): Temperature as a function of position for the TI heterostructure setup. The parameter is $M_3/m=3+0.2x/\xi$.}\label{Fig:demonstrate}
\end{figure}

\emph{Experimental detection--} 
Here we discuss how our prediction linking the temperature profile to Weyl node separation, Eq.~\eqref{Eq:Temperature}, can be detected in topological insulator-magnetic insulator layer structure realization of Weyl semimetals~\cite{PhysRevLett.107.127205,PhysRevB.84.235126}. In this case the magnetization perpendicular to the topological insulator layers is mapped to $M_3$ and our theory is readily applicable. The experimental setup is shown in Fig.~\ref{Fig:demonstrate}~(b). The inhomogeneous magnetization is realized by applying an decreasing magnetic field $B_z$ along the $x$ direction. There is no need for a high precision control over the field gradient as long as there is detectable increase in magnetization over the sample.  When the temperature at $x=0$ is fixed by a heat bath, there will be a temperature gradient along the $x$ direction. Fig.~\ref{Fig:demonstrate}~(c) shows the result of the temperature as a function of position for $M_3/m=3+0.2x/\xi$. Alternatively, one could employ a naturally occurring N\'eel-type texture by contacting the sample to the heat bath near a domain wall and study the temperature profile in the vicinity.

Other candidate materials to observe this effect include ferromagnetic Weyl semimetals which have been recently identified experimentally~\cite{Belopolski1278,Liu1282,Morali1286,Wang_2019,Soh_2019,Su_2019,Ma_2019} and are also captured by our theory. Specifically, we consider EuCd$_2$As$_2$, whose ground state is an itinerant magnet where Eu ions form ferromagnetic layers stacked antiferromagnetically along the $c$-axis. When a magnetic field  is applied along the $c$-axis, EuCd$_2$As$_2$ becomes a Weyl semimetal with only two Weyl points~\cite{Wang_2019,Soh_2019}. Above the N\'eel temperature, the Weyl nodes have been observed~\cite{Ma_2019} even in the absence of external magnetic field, indicating strong ferromagnetic correlations.
The bands near the Fermi level are dominated by the Cd 5$s$ and As 4$p$ orbitals. A theoretical description in terms of $4\times 4$ $\mathbf{k}\cdot\mathbf{p}$ Hamiltonian for the ferromagnetic state around the $\Gamma$ point has been constructed in Ref.~\cite{Wang_2019}. The four band model can be further mapped to Eq.~\eqref{Eq:Weyl} with  $d_1=A k_x$, $d_2=A k_y$, and $d_3=M_3+m-C_1 k^2_z$ where  $m$ is the gap at $\Gamma$ point and $A$ and $C_1$ are material parameters (see the SI for details). The magnetization energy difference $M_3$ between the $p$ and $s$ orbitals  arises from the exchange couplings between Eu magnetic moments and Cd and As orbitals. Thus, $M_3$ can be tuned by the spin polarization of Eu irons through external magnetic field. Therefore, a position dependent magnetic field can induce spatially varying $M_3$, which in turn leads to a temperature gradient through Eq.~\eqref{Eq:Temperature}. Magnetization will directly couple to electrons so the temperature profile~\eqref{Eq:Temperature} refers to electron temperature. When electron-phonon mediated relaxation processes cannot be neglected, the predicted temperature profile will relax. However, at low-temperatures the relaxation time becomes long and the temperature gradient can be observed after a bath temperature quench.

\emph{Conclusion and outlook-- } In summary, we developed an effective theory for magnetic Weyl semimetals where inhomogeneous magnetizations are mapped to spacetime torsion as well as axial gauge fields. Building on this notion, we discovered a novel topological response, the thermal magnetorosional effect, which implies that an inhomogeneous magnetization will induce a temperature gradient in Weyl semimetals. This intriguing effect is rooted in the mixed chiral-torsional anomaly at finite temperature,  which unlike the zero temperature counterpart~\cite{Chandia_1997,Onkar_2014} is cutoff independent, and therefore is of topological nature. We demonstrated how to detect this effect experimentally in recently discovered ferromagentic Weyl semimetals  as well as a TI-magnet heterostructure. Our discovery would be used as magnetic gradient refrigeration, and combining it with other thermoelectric effects provides new methods to control responses of Weyl semimetals. 
Our prediction the TME also provides an alternative and concrete way to experimentally verify the torsion anomaly~\cite{PhysRevLett.122.056601} which has stirred controversy in the field theory literature~\cite{Mielke_1999,PhysRevD.63.048501,PhysRevD.63.048502}.

In the present work we treated magnetization as static field. Allowing for dynamical effects, our results imply that it is possible to manipulate magnetization by using temperature gradients. Furthermore, under those circumstances we expect exotic collective magnetic excitations resulting from the mixed chiral-torsional anomaly. To study these effects, the dynamics of the magnetization must be included in the effective theory which we leave for future research. 
Another interesting issue for future study is to develop a theory of elastic analogue of the TME. Since our effective theory for Weyl fermions is analogous to the one obtained in the presence of lattice deformations~\cite{Cortijo_2016},  we expect that also strain would give rise to thermal effects. Torsion also emerges in topological superfluids and superconductors~\cite{Golan_2018,NissinenVolovik_2019,nissinen2019emergent}, where the present theory may not be directly applicable because the central charge of the edge mode in that case~\cite{ReadGreen_2000} is different from the present model and the anomaly might be different. This presents another interesting avenue for future work.

\emph{Acknowledgment--} The authors acknowledge Aalto Center for Quantum Engineering for financial support. L.L. acknowledges useful discussions with Tao Jiang.

\bibliography{MagnetothermalWeyl.bib}

\end{document}


\widetext

\begin{center}
	\Large SUPPLEMENTAL INFORMATION to Topological magnetotorsional effect in Weyl semimetals
\end{center}

\section{Effective action for Weyl fermions}
To obtain the low energy linearized Weyl Hamiltonian from the four-band parent Hamiltonian, we first rotate the $\mathbf{M}\cdot \mathbf{b}$ along the $z$ direction by using a unitary transformation $W(\mathbf{r})=e^{-i\frac{\phi}{2}b_3}e^{-i\frac{\theta}{2}b_2}$, and this leads to
\begin{eqnarray}
H'&=&W^\dag H W=\frac{1}{2}\{k_i -\omega_i, E^i_{~a}\gamma^a\} +m\gamma_4+M(\mathbf{r})\gamma_{12},\\
&=& \kappa_i \gamma_i+m\gamma_4+ M \gamma_{12} +f \gamma_{45},
\end{eqnarray}
where 
$W^\dag \gamma_i W=E^i_{~a} \gamma_a$, $\omega_i(\mathbf{r})=i W^\dag(\mathbf{r}) \partial_{r_i} W(\mathbf{r})\equiv \omega_{i}^{~a}b_a$, $f=E^i_{~a}\omega_i^{~a}$, 
$\kappa_1=\cos{\theta}\cos{\phi} k_x+\cos{\theta}\sin{\phi} k_y-\sin{\theta}k_z$,
$\kappa_2=-\sin{\phi}k_x+\cos{\phi}k_y$,
$\kappa_3=\cos{\phi}\sin{\theta}k_x+\sin{\phi}\sin{\theta}k_y+\cos{\theta}k_z$. (The expressions for $\kappa_i$ should be understood as symmetrized.) We then use a Schrieffer-Wolff transformation to arrive at the effective two band Weyl Hamiltonian
\begin{eqnarray}
H_W&=& d_a(\mathbf{k},\mathbf{r})\sigma^a= \sum_{i=1,2} \kappa_i \sigma_i +\left(-m+M(\mathbf{r})-\frac{\kappa_3^2 +f^2 }{2m}\right)\sigma_z-\left(\frac{f \kappa_3 }{m} \right) \sigma^0.
\end{eqnarray}
The Weyl points are determined by treating $\mathbf{k}$ and $\mathbf{r}$ as $c$-numbers and solving $d_1=d_2=d_3=0$.
We find two Weyl points given by $\pm \mathbf{K}_W=\pm K_W(\sin{\theta}\cos{\phi}, \sin{\theta}\sin{\phi},\cos{\theta})$ with $K_W=\sqrt{2m(M-m)-f^2}$. Expanding the effective Hamiltonian around $\mathbf{K}_W$, we arrive at the low energy Hamiltonian for the left-handed Weyl fermions, 
\begin{eqnarray}
H_L \approx \frac{1}{2}\{v^i_{~a} ,k_i-K_{W,i}\}\sigma^a - \frac{fK_W}{m}\sigma^0,
\end{eqnarray}
where 
$v^i_{~a}(\mathbf{r})=\frac{\partial d_a}{\partial k_i}\big|_{\mathbf{K}_W}$. It is convenient to introduce $v^{0}_{~0}=-1$ and $v^{0}_{~1,2,3}=0$, then
\begin{eqnarray}
v^\mu_{~a}=\left[\begin{array}{cccc}
-1 & 0 & 0 & 0 \\ 
-\frac{f}{m}\cos{\phi}\sin{\theta} & \cos{\theta}\cos{\phi} & -\sin{\phi} & -\frac{K_W}{m}\cos{\phi}\sin{\theta}\\ 
-\frac{f}{m}\sin{\phi}\sin{\theta} & \cos{\theta}\sin{\phi} & \cos{\phi}  & -\frac{K_W}{m}\sin{\phi}\sin{\theta}\\
-\frac{f}{m}\cos{\theta} & -\sin{\theta} & 0 & -\frac{K_W}{m}\cos{\theta}
\end{array} \right],
\end{eqnarray}
and
\begin{eqnarray}
v_\mu^{~a}=\left[\begin{array}{cccc}
-1 & 0 & 0 & f/K_W \\ 
0 & \cos{\theta}\cos{\phi} & -\sin{\phi} & -\frac{m}{K_W}\cos{\phi}\sin{\theta}\\ 
0 & \cos{\theta}\sin{\phi} & \cos{\phi}  & -\frac{m}{K_W}\sin{\phi}\sin{\theta}\\
0 & -\sin{\theta} & 0 & -\frac{m}{K_W}\cos{\theta}
\end{array} \right].
\end{eqnarray}

The action for the left-handed Weyl fermions reads
\begin{eqnarray}
S&=&\int\mathrm{d}^4x~\psi^\dag_L (i\partial_t-H_L)\psi_L,\\
&=&\int\mathrm{d}^4x~\frac{1}{2}\psi^\dag_L v^{\mu}_{~a}(i\partial_\mu+A_{5,\mu}) \sigma^a\psi_L+\mathrm{h.c.},
\end{eqnarray}
where $\psi_L$ is a two component spinor and $A_{5}=(fK_W/m, \mathbf{K}_W)$ is the effective gauge field. 
To cast the action into the standard form in curved spacetime, we introduce new frame field $e_\mu^{~a}$ such that $\mathrm{det}(e_\mu^{~a}) e^{\mu}_{~a}=v^\mu_{~a}$, then $\mathrm{det}(e_\mu^{~a})= (\mathrm{det} v^\mu_{~a})^{1/3}\equiv e=\sqrt{-\mathrm{det}g_{\mu\nu}}=(K_W/m)^{1/3}$ and
\begin{eqnarray}
S
&=&\int\mathrm{d}^4x~\frac{e}{2}\psi^\dag_L e^{\mu}_{~a}(i\partial_\mu+A_\mu) \sigma^a\psi_L+\mathrm{h.c.},\\
&=&\int\mathrm{d}^4x~e\psi^\dag_L e^{\mu}_{~a}\left(i\partial_\mu+A_\mu+i \frac{T_{\mu}}{2}\right) \sigma^a\psi_L.
\end{eqnarray}
The above action describes a Weyl fermion in a spacetime without curvature (because the spin connection is zero) but with torsion, $T^a=d e^a$. The trace of the torsion tensor is $T_{\mu}=(\partial_{\mu} e_{\nu}^{~b}-\partial_{\nu} e_{\mu}^{~b})e^\nu_{~b}$. Using $\partial_\mu e = e e^\nu_{~a}\partial_\mu e_{\nu}^{~a}$, we get $e e^{\mu}_{~a} T_{\mu}=\partial_{\nu}(e e^\nu_{~a})$.

The action is invariant under the Nieh-Yan-Weyl transformation~\cite{Hughes_2013},
\begin{eqnarray}
&& e^a\to \exp{[\Lambda(x)]}e^a,\\
&& \psi\to \exp{[-(d-1)\Lambda(x)/2]}\psi.
\end{eqnarray}
Note that the gauge field is unchanged. This symmetry is broken by the mixed chiral Nieh-Yan anomaly since the anomaly action is not invariant under the transformation. 

\section{Direct calculations of the axial density}
In this section we calculate the axial density directly from the torsional Landau levels. This tells us how to determine the normalization of the volume element. 
We consider the following Hamiltonian with anisotropic Fermi velocity
\begin{eqnarray}
H&=&v_xk_x \sigma_x+(v_yk_y -T^3_B x v k_z)\sigma_y+v_zk_z\sigma_z,\\
&=&\left[\begin{array}{cc}
v_z k_z& v_x k_x-i (v_yk_y -T^3_B x v k_z) \\ 
v_x k_x+i(v_yk_y -T^3_B x v k_z) & -v_z k_z
\end{array}\right],\\
&=&\left[\begin{array}{cc}
v_z k_z& v_x k_x+i v T^3_B x k_z -iv_y k_y \\ 
v_x k_x-i v T^3_B x k_z +iv_y k_y & -v_z k_z
\end{array}\right].
\end{eqnarray}
We assume $v_i>0$ and $T^3_B>0$. The dimension of $k_i$ and $T^3_B$ are Length$^{-1}$ and the dimension of $v_i$ is Energy$\cdot$Length.
Introducing $E^2_B=v_xvT^3_B|k_z|$ and  the bosonic annihilation operator $a$, then   
for $k_z>0$, the Hamiltonian can be written as
\begin{eqnarray}
H
&=&\left[\begin{array}{cc}
v_z k_z& \sqrt{2}E_B a^\dag \\ 
\sqrt{2}E_B a & -v_z k_z
\end{array}\right],
\end{eqnarray}
and for $k_z<0$
\begin{eqnarray}
H
&=&\left[\begin{array}{cc}
v_z k_z& \sqrt{2}E_B a \\ 
\sqrt{2}E_B a^\dag & -v_z k_z
\end{array}\right].
\end{eqnarray}
So the energy dispersion is
\begin{eqnarray}
E_n=\bigg\{  \begin{array}{c}
v_z|k_z|, n=0,\\ 
\mathrm{sgn}(n) \sqrt{v^2_zk^2_z+2|n| E^2_B}, |n|>1.
\end{array} 
\end{eqnarray}  
The degeneracy of the torsional Landau levels is 
\begin{eqnarray}
\frac{1}{2\pi}\frac{v T^3_B |k_z|}{v_y}.
\end{eqnarray}  
The axial charge density is~\cite{Huang_2019}
\begin{eqnarray}
j^0_5=\frac{T^2}{12}\frac{v T^3_B}{v_y v^2_z},
\end{eqnarray}
where we only write down the cutoff independent part. 
The velocity matrix corresponding to the above Hamiltonian is
\begin{eqnarray}
v^{i}_{~a}=\left[\begin{array}{ccc}
v_x & 0 &0  \\ 
0 & v_y & 0 \\ 
0 & -T^3_B x v & v_z
\end{array} \right],
\end{eqnarray} 
and its inverse is 
\begin{eqnarray}
v_{i}^{~a}=\left[\begin{array}{ccc}
1/v_x & 0 &0  \\ 
0 & 1/v_y & T^3_B v x/ v_y v_z \\ 
0 & 0 & 1/v_z
\end{array} \right],
\end{eqnarray} 
so the chiral density can be written as
\begin{eqnarray}\label{Eq:J5_qm}
j^0_5=\frac{T^2}{12}v_z^{~3}\partial_x v_{y3}.
\end{eqnarray}

Now let us rewrite the above result using the frame fields.
We first write down the action
\begin{eqnarray}\label{Eq:Action}
\int d^4x~ c \psi^\dag \left(i \partial_{ct} - \frac{v^i_{~a}}{ c} k_i\sigma^a\right)\psi,
\end{eqnarray}
where $x_0=ct$ and $c$ is a parameter to be determined.
The velocity matrix becomes
\begin{eqnarray}
v^{\mu}_{~a}=\left[\begin{array}{cccc}
-1& 0 & 0& 0    \\
0 & v_x/c & 0 &0  \\ 
0 & 0 & v_y/c & 0 \\ 
0 & 0 & -T^3_B x v/c & v_z/c
\end{array} \right],
\end{eqnarray} 
and the coframe fields are
\begin{eqnarray}
e_{\mu}^{~a}=e\left[\begin{array}{cccc}
-1& 0 & 0& 0    \\
0 & c/v_x & 0 &0  \\ 
0 & 0 & c/v_y & cT^3_B v x/ v_y v_z \\ 
0 & 0 & 0 & c/v_z
\end{array} \right],
\end{eqnarray}
with $e^3=v_x v_y v_z/c^3$. 

The axial current obtained from the anomaly action reads [the prefactor $1/c^2$ is because of the overall $c$ in Eq.~\eqref{Eq:Action}]
\begin{eqnarray}
e j^\mu_5=\frac{T^2}{12 c^2}\varepsilon^{\mu\nu\rho\sigma}e_\nu^{~a}\partial_\rho e_{\sigma a}, 
\end{eqnarray}
from which we get
\begin{eqnarray}\label{Eq:J5_qft}
e j^0_5 =\frac{T^2}{12}\frac{e^2}{c^2}\frac{c^2 T^3_B v}{v_y v^2_z}.
\end{eqnarray}
Matching Eq.~\eqref{Eq:J5_qft} and Eq.~\eqref{Eq:J5_qm},  the velocity parameter is fixed to be $c^3=v_x v_y v_z$, i.e., we should use the normalization $e=1$ in the action Eq.~\eqref{Eq:Action}. Then we always have $e_{0}^{~0}=1$, and  as a result,  the torsional electric field vanishes if the frame fields are independent of time.

\section{Topological magnetotorsional effect for rotating magnetization}
In the main text we discussed the topological magnetotorsional effect for unidirectional magnetization. Here we study a N\'eel type rotating magnetic texture and show that the effect is described by a similar equation as Eq.~\eqref{Eq:Temperature}  in the main text.
When the  magnetization rotates in the $x-z$ plane,
\begin{eqnarray}
\mathbf{M}(x)=M(\sin{(\theta(x))},0,\cos(\theta(x))),
\end{eqnarray}
with $M$ being a constant,
then the axial `gauge' field is
\begin{eqnarray}
\mathbf{A}_5=K_W (\sin{(\theta(x))},0,\cos(\theta(x))),
\end{eqnarray}
with $K_W=\sqrt{2m(M-m)}$. Since the magnetization induced torsional electric field vanishes, the 
the temperature gradient is determined by Eq.~\eqref{TME} in the main text
\begin{eqnarray}
\frac{T}{6}\mathbf{A}_5\times\nabla T-\frac{T^2}{12}\nabla\times \mathbf{A}_5=0,
\end{eqnarray}
which gives
\begin{eqnarray}
&&-\frac{T}{6}A_{5,z}\partial_y T=0,\label{Eq:TME_diffy}\\
&&\frac{T}{6}(A_{5,z}\partial_x T-A_{5,x} \partial_z T)=-\frac{T^2}{12}\partial_x A_{5,z}.\label{Eq:TME_diffx}
\end{eqnarray}
The temperature at $x=0$ plane is fixed by the thermal bath, and the temperature profile is
\begin{eqnarray}
\frac{T(x)}{T(0)}=\sqrt{\frac{K_{W,z}(0)}{K_{W,z}(x)}}.
\end{eqnarray}

Similar result can be obtained for the N\'eel texture where the magnetization rotates in the $x-y$ plane,
\begin{eqnarray}
\mathbf{M}(x)=M(\sin{(\theta(x))},\cos(\theta(x)),0),
\end{eqnarray}
then the corresponding differential equations are
\begin{eqnarray}
&&\frac{T}{6} A_{5,y} \partial_z T=0,\\
&&\frac{T}{6} A_{5,z} \partial_z T=0,\\
&& \frac{T}{6}(A_{5,x}\partial_y T-A_{5,y}\partial_x T)=\frac{T^2}{12} \partial_x A_{5,y}.
\end{eqnarray}
Then the temperature profile is
\begin{eqnarray}
\frac{T(x)}{T(0)}=\sqrt{\frac{K_{W,y}(0)}{K_{W,y}(x)}}.
\end{eqnarray}
Combining the results from the Neel textures on orthogonal planes, we obtain 
\begin{eqnarray}
\frac{T(x)}{T(0)}=\sqrt{\frac{K_{W\perp}(0)}{K_{W\perp}(x)}},
\end{eqnarray}
where $k_{W\perp}$ denotes the projection of the Weyl node  which is perpendicular to the temperature gradient.

\section{$\mathbf{k}\cdot \mathbf{p}$ model for $\mathbf{EuCd}_2\mathbf{As}_2$}
The bands near the Fermi level are dominated by the As anti-bonding $p$ orbitals $|p,\pm\frac{3}{2}\rangle^-$ and Cd bonding $s$ orbitals $|s,\pm\frac{1}{2}\rangle^+$, and 
the effective $4\times 4$ Hamiltonian near the $\Gamma$ point takes the form~\cite{PhysRevB.98.201116,Wang_2019} (in the basis $|s,\frac{1}{2}\rangle^+$, $|p,\frac{3}{2}\rangle^-$,$|s,-\frac{1}{2}\rangle^+$,$|p,-\frac{3}{2}\rangle^-$)
\begin{eqnarray}
H=\epsilon(\mathbf{k})+\left[\begin{array}{cccc}
m-C_1k^2_z-C_2k^2_{||} & Ak_+ & 0 & Dk_+ \\ 
A k_- & -(m-C_1k^2_z-C_2k^2_{||} ) & Dk_+ & 0\\ 
0 & Dk_- & m-C_1k^2_z-C_2k^2_{||}   & -A k_- \\
Dk_- & 0& -A k_+  & -(m-C_1k^2_z-C_2k^2_{||} )
\end{array} \right],
\end{eqnarray}
where $\epsilon(\mathbf{k})=m_0+m_1k^2_z+m_2k^2_{||}$, $k_\pm=k_x\pm i k_y$ and $k^2_{||}=k^2_x+k^2_y$. The parameters $m_0$, $C_1$, and $C_2$ are negative such that there is band inversion. 
In the ferromagnetic state, Eu irons are polarized by external magnetic field, and due to the exchange interaction between Eu and As and Cd, there will be extra energy spitting between the four orbitals. We do not know the details of exchange interaction, so we use a Zeeman term to split the energy levels, although the Zeeman splitting is much smaller comparing to the exchange energy. The Zeeman term can be written as~\cite{Cano_2017,Wang_2019}
\begin{eqnarray}
H_Z=\left[\begin{array}{cccc}
\frac{g_s}{2} B & 0 & 0 & 0 \\ 
0 & \frac{3g_p}{2} B & 0 & 0\\ 
0 & 0 & -\frac{g_s}{2} B   &0 \\
0& 0&0  & -\frac{3g_p}{2}
\end{array} \right].
\end{eqnarray}
The Zeeman term breaks the two fold degeneracy and leads to Weyl points. Suppose the band crossing occurs between $|p,\frac{3}{2}\rangle^-$, $|s,-\frac{1}{2}\rangle^+$ states, the the effective Weyl Hamiltonian can be written as
\begin{eqnarray}
H_W=\epsilon(\mathbf{k})+\left[\begin{array}{cc}
-m+C_1k^2_z+M & Dk_+  \\ 
D k_- & m-C_1k^2_z-M
\end{array} \right],
\end{eqnarray}
where $M=\frac{3g_p}{4}B+\frac{g_s}{4}B$ and the $k^2_{||}$ has been discarded. The band crossing may occur between other bands, however, the effective Weyl Hamiltonian takes the same form. 
Expanding around the Weyl point $\mathbf{K}_W=(0,0,k_c)$, we get the low energy Weyl Hamiltonian
\begin{eqnarray}
H_W\approx v^z_{~0}k_z\sigma^0+v^x_{~1}k_x\sigma^1 +v^y_{~2}k_y\sigma^2+v^z_{~3}k_z\sigma^3.
\end{eqnarray}
The parameters in the fully polarized phase are~\cite{Soh_2019} $v^z_{~0}=0.16$eV$\cdot$nm, $v^z_{~3}=0.2$eV$\cdot$nm, and $v^x_{~1}=v^y_{~2}=0.3$eV$\cdot$nm. It can be seen from the effective model that $v^z_{~0}$ and $v^z_{~3}$ can be tuned by external magnetic field. 
For this simple example, the magnetization induced  torsional electric field vanishes, so in equilibrium 
\begin{eqnarray}
\frac{T}{6}\mathbf{A}_5\times\nabla T-\frac{T^2}{12}\nabla\times \mathbf{A}_5=0.
\end{eqnarray}
Substituting $\mathbf{A}_5=(0,0,K_W(x))$ into the above equation, we find
\begin{eqnarray}
\frac{T(x)}{T(0)}=\sqrt{\frac{K_W(0)}{K_W(x)}}.
\end{eqnarray}

\bibliography{MagnetothermalWeyl.bib}